\documentstyle[12pt]{article}

\catcode`\@=11

\@addtoreset{equation}{section}
\@addtoreset{footnote}{section}
\@addtoreset{footnote}{subsection}
\catcode`\@=12

\newtheorem{theorem}{Theorem}
\newenvironment{acknowledgements}%
  {\bigskip\noindent{\bf Acknowledgements}\bigskip}%
  {\bigskip}
\newenvironment{noteadded}%
  {\bigskip\noindent{\bf Note added}\bigskip}%
  {\bigskip}
\newcommand{\beqn}{\begin{equation}}
\newcommand{\eeqn}{\end{equation}}
\newcommand{\beqnarray}{\begin{eqnarray}}
\newcommand{\eeqnarray}{\end{eqnarray}}
\newcommand{\rd}{\partial}
\newcommand{\dfrac}[2]{ \frac{\displaystyle #1}{\displaystyle #2} }

\newcommand{\diag}{\mathop{\mbox{\rm diag}}}
\newcommand{\Tr}{\mathop{\mbox{\rm Tr}}}
\newcommand{\const}{\mbox{\rm const.}\;}

\newcommand{\Abar}{{\bar{A}}}

\newcommand{\Bbar}{{\bar{B}}}

\newcommand{\cbar}{{\bar{c}}}
\newcommand{\Cbar}{{\bar{C}}}

\newcommand{\bL}{{\bf L}}
\newcommand{\Lbar}{{\bar{L}}}

\newcommand{\bLbar}{{\bar{\bL}}}

\newcommand{\bM}{{\bf M}}
\newcommand{\Mbar}{{\bar{M}}}

\newcommand{\bMbar}{{\bar{\bM}}}
\newcommand{\bU}{{\bf U}}

\newcommand{\bW}{{\bf W}}
\newcommand{\bZ}{{\bf Z}}
\newcommand{\Wbar}{{\bar{W}}}
\newcommand{\Xbar}{{\bar{X}}}

\newcommand{\bWbar}{{\bar{\bW}}}
\newcommand{\abar}{{\bar{a}}}

\newcommand{\pbar}{{\bar{p}}}
\newcommand{\tbar}{{\bar{t}}}
\newcommand{\ubar}{{\bar{u}}}
\newcommand{\utilde}{\tilde{u}}
\newcommand{\vbar}{{\bar{v}}}
\newcommand{\wbar}{{\bar{w}}}

\newcommand{\bLambda}{{\bf \Lambda}}
\newcommand{\bDelta}{{\bf \Delta}}

\begin{document}

\begin{flushright}
\baselineskip=12pt
KUCP-0079\\
hep-th/9506089\\
June 1995
\end{flushright}

\begin{center}
\LARGE
    Toda Lattice Hierarchy and\\
    Generalized String Equations\\
\bigskip
\bigskip
\Large
    Kanehisa Takasaki
    \footnote{E-mail: takasaki@yukawa.kyoto-u.ac.jp}\\
\bigskip
\normalsize\it
    Department of Fundamental Sciences\\
    Faculty of Integrated Human Studies, Kyoto University\\
    Yoshida, Sakyo-ku, Kyoto 606, Japan\\
\end{center}
\bigskip
\bigskip
\bigskip
\begin{abstract}
\noindent
String equations of the $p$-th generalized Kontsevich model and
the compactified $c = 1$ string theory are re-examined in the
language of the Toda lattice hierarchy.  As opposed to a hypothesis
postulated in the literature, the generalized Kontsevich model at
$p = -1$ does not coincide with the $c = 1$ string theory at
self-dual radius. A broader family of solutions of the Toda
lattice hierarchy including these models are constructed, and
shown to satisfy generalized string equations.  The status of a
variety of $c \le 1$ string models is discussed in this new
framework.
\end{abstract}
\newpage


\section{Introduction}

The so called ``string equations'' play a key role in
various applications of integrable hierarchies to
low dimensional string theories. The most fundamental
integrable hierarchy in this context is the KP hierarchy
\cite{bib:KP} that provides a universal framework for
dealing with many KdV-type hierarchies. String equations
for ``$(p,q)$ models'' of two-dimensional quantum gravity
can be treated in a unified manner in this language.
In contrast, the status of the Toda lattice hierarchy
\cite{bib:UTToda}, which is another universal integrable
hierarchy, had remained relatively obscure until rather
recent years. The Toda lattice hierarchy was pointed out
to be an integrable structure of the one- and multi-matrix
models \cite{bib:MMMToda}, but these matrix models
(matrix integrals) were only considered as an intermediate
step towards the continuous (double scaling) limit to
two-dimensional quantum gravity.

In the last few years, the Toda lattice hierarchy has come
to be studied from renewed points of view, such as $c = 1$
strings \cite{bib:DMP,bib:Marshakov,bib:Nakatsu,bib:NTTceq1},
two-dimensional topological strings \cite{bib:GhoshalMukhi,%
bib:HOP,bib:EK2Dstring,bib:T2Dstring,bib:BX2Dstring},
the topological $CP^1$ sigma model and its variations
related to affine Coxeter groups
\cite{bib:EguchiYang,bib:KannoOhta,bib:EguchiHoriYang}.
As opposed to the $(p,q)$ models in the KP hierarchy,
these are related to string theories with a true continuous
target space. Our goal in this paper is to elucidate the
structure of those string equations, in particular, those
of $c = 1$ strings in a more general framework.

It will be instructive to recall the relationship
between the $(p,q)$ models and the KP hierarchy.
String equations of these models were first discovered
in the form of the Douglas equation \cite{bib:Douglas}
\beqn
    [ P, Q ] = 1,
\eeqn
where $P$ and $Q$ are ordinary differential operators
of the form
\beqnarray
    P &=& \rd_x^p + a_2 \rd_x^{p-2} + \cdots + a_p,
                                     \nonumber \\
    Q &=& \rd_x^q + b_2 \rd_x^{q-2} + \cdots + b_q.
\eeqnarray
These two operators were later found to be related
to an extended Lax formalism of the KP hierarchy.
The extended Lax formalism, developed by Orlov
and his coworkers \cite{bib:Orlovetal},  is based
on the ordinary Lax operator $L$ and a secondary
Lax operator (Orlov-Shulman operator) $M$. They
are pseudo-differential operators of the form
\beqnarray
    L &=& \rd_x + \sum_{n=1}^\infty u_{n+1} \rd_x^{-n},
                                       \nonumber \\
    M &=& \sum_{n=2}^\infty n t_n L^{n-1} + x
          + \sum_{n=1} v_n L^{-n-1},
\eeqnarray
and obey the Lax equations
\beqn
     \dfrac{\rd L}{\rd t_n} = [B_n,L], \quad
     \dfrac{\rd M}{\rd t_n} = [B_n,M]
\eeqn
and the canonical commutation relation
\beqn
    [L,M] = 1.
\eeqn
The operators $B_n$ are given by
\beqn
    B_n = ( L^n )_{\ge 0},
\eeqn
where ``$(\quad)_{\ge 0}$'' denotes the projection
onto the space spanned by nonnegative powers of
$\rd_x$; similarly, we shall use ``$(\quad)_{\le -1}$''
for the projection onto the space spanned by negative
powers of $\rd_x$. To reproduce the Douglas equation,
we define $P$ and $Q$ as
\beqnarray
    P &=& L^p,
                               \nonumber \\
    Q &=& - \frac{1}{p} M L^{1-p}
          + \frac{p-1}{2p} L^{-p} + L^q,
\eeqnarray
require the constraints
\beqn
    ( P )_{\le -1} = 0, \quad
    ( Q )_{\le -1} = 0,
                        \label{eq:PQconstraint}
\eeqn
and restrict the range of the time
variables as
\beqn
    t_{p+q} = t_{p+q+1} = \cdots = 0.
                        \label{eq:smallPS}
\eeqn
The Douglas equation follows automatically from the
construction of $P$ and $Q$ --- (\ref{eq:PQconstraint})
and (\ref{eq:smallPS}) ensure that $P$ and $Q$ are
differential operators of the required form.  Note,
in particular, that (\ref{eq:PQconstraint}) give
constraints on the Lax and Orlov-Shulman operators,
the first one being a reduction condition to
the $p$-th generalized KdV hierarchy.
Thus the $(p,q)$ model can be understood as
a ``constrained KP hierarchy''. From the standpoint
of the KP hierarchy, therefore, it is these constraints
on $L$ and $M$ rather than the Douglas equation itself
that plays a more essential role.  In our terminology,
``string equations'' mean such constraints as
(\ref{eq:PQconstraint}).

This correspondence with the KP hierarchy allows one
to use many powerful tools developed for the study
of the KP hierarchy, such as the Sato Grassmannian,
the Hirota equations, $W_{1+\infty}$ algebras, etc.
\cite{bib:KacSchwarz,bib:Goeree,bib:Yoneya,bib:FKN}.
Since this raises many interesting mathematical issues,
generalizations to multi-component KP hierarchies
have been also attempted \cite{bib:vandeLeur}.

The case of $(p,1)$ (or $(1,q)$) models has been
studied with particular interest in recent studies.
This is the case where the models becomes
``topological'', i.e., describes topological
strings of $A_k$ ($k = p - 2$) type \cite{bib:Dijkgraaf}.
Unlike the other $(p,q)$ models,  string equation for
these models can be solved more explicitly in terms of
a ``matrix Airy function'' (or ``Kontsevich integral'')
\cite{bib:AvM}, and by virtue of this matrix integral
construction, it is rigorously proven (at least for
the case of $p = 2$) that the corresponding $\tau$
function is a generating function of intersection
numbers on a moduli space of Riemann surfaces
\cite{bib:Kontsevich}. The (generalized) Kontsevich
integral is an integral over the space of $N \times N$
Hermitian matrices of the form
\beqn
    Z(\Lambda) = C(\Lambda) \int dM
      \exp \Tr( \Lambda^p M - \frac{1}{p+1} M^{p+1}),
\eeqn
where $\Lambda$ is another $N \times N$ Hermitian
matrix, and $C(\Lambda)$ is a normalization factor
that also plays an important role.  Kontsevich's
observation is that this integral has two-fold
interpretations.  The first interpretation,
revealed by  ``fat graph'' expansion, is that
this is a generating function of intersection
numbers.  The second is that this is a $\tau$ function
of the KP hierarchy,
\beqn
    Z(\Lambda) = \tau(t),
\eeqn
where $\Lambda$ and $t$ are connected by the so called
``Miwa transformation'':
\beqn
    t_n = \frac{1}{n} \Tr \Lambda^{-n}
        = \sum_{i=1}^N \lambda_i^{-n}.
\eeqn
Here $\lambda_1, \ldots, \lambda_N$ are eigenvalues of
$\Lambda$.

As for the Toda lattice hierarchy, our knowledge
on string equations is far more fragmental, but
simultaneously suggests richer possibilities.
The most fundamental and well understood cases
are the one- and two-matrix models
\cite{bib:GKMvsToda,bib:BXMMM,bib:Morozov}.
String equations of these two models are quite different.
Let us specify this in more detail. As we shall review
in the next section, the Lax formalism of the
Toda lattice hierarchy uses two Lax operators
$L,\Lbar$ and two Orlov-Shulman operators $M,\Mbar$
\cite{bib:TTToda,bib:ASvM}.
These operators are ``difference'' operators
obeying a set of Lax equations and a twisted
canonical commutation relations of the form
\beqn
    [L, M] = L, \quad [\Lbar,\Mbar] = \Lbar.
\eeqn
($M$ should not be confused with the matrix variable
$M$ in the Kontsevich integral.) String equations are
formulated as algebraic relations between the two
pairs $(L,M)$ and $(\Lbar,\Mbar)$.  Roughly speaking,
string equations of the one-matrix model are written
\beqn
    L = \Lbar^{-1}, \quad M L^{-1} = - \Mbar \Lbar,
                             \label{eq:1MMStringEq}
\eeqn
and those of the two-matrix model are given by
\beqn
    L = - \Mbar \Lbar, \quad M L^{-1} = - \Lbar^{-1}.
                             \label{eq:2MMStringEq}
\eeqn
The point is that both $(L,ML^{-1})$ and
($\Lbar^{-1},\Mbar \Lbar)$ are a canonical conjugate
pair, and that string equations are canonical
transformations between them.  The difference between
the one- and two-matrix models is that whereas string
equations (\ref{eq:1MMStringEq}) of the one-matrix model
are coordinate-to-coordinate and momentum-to-momentum
relations, string equations (\ref{eq:2MMStringEq})
of the two-matrix model mix coordinate and momentum
variables.  Note that the latter is also a
characteristic of Fourier transformations. This is
actually related to the fact that the Kontsevich
integral is essentially a matrix version of Fourier
transformations \cite{bib:Morozov}.

In fact, most examples of string equations in the Toda
lattice hierarchy are variants of the above two.
For instance, string equations of $c = 1$ strings
\cite{bib:DMP,bib:Marshakov,bib:Nakatsu}
and their topological versions
\cite{bib:GhoshalMukhi,bib:HOP,bib:EK2Dstring,%
bib:T2Dstring,bib:BX2Dstring}
are of the two-matrix model type. The generalized
Kontsevich models, too, may be considered as a solution
of the Toda lattice hierarchy obeying string equations of
the two-matrix model type \cite{bib:GKMvsToda}.
Meanwhile, string equations of the topological
$CP^1$ model and its variants
\cite{bib:EguchiYang,bib:KannoOhta,bib:EguchiHoriYang}
are of the one-matrix type. Furthermore, the deformed
$c = 1$ theory in the presence of black hole backgrounds
\cite{bib:NTTceq1} are known to obey more involved string
equations, though this case, too, is essentially of the
two-matrix model type.

In this paper, we are mostly concerned with
string equations of the two-matrix model type
in the above sense. We re-examine the
generalized Kontsevich models and the
$c = 1$ string theory in detail, and
present a broader family of solutions that
includes these two examples as special cases.
This will clarify the status of a variety of
$c \le 1$ string models. For instance, we shall
show that the generalized Kontsevich model
at $p = -1$ does not reproduce the $c = 1$
string theory, but is rather related to the
Penner model \cite{bib:Penner}. This poses
a question on recent attempts in the literature
\cite{bib:Ozetal} that treat two-dimensional
topological strings as ``$A_k$ strings at $k = -3$''.

This paper is organized as follows.  Section 2 is
a brief review on necessary tools and results from
the theory of the Toda lattice hierarchy. Sections
3 and 4 deal with the generalized Kontsevich models
and the $c = 1$ string theory. In Section 5, our new
family of solutions and string equations are presented.
Section 6 is devoted to conclusion and discussion.

\section{Preliminaries on Toda lattice hierarchy}

We first present necessary tools and results \cite{bib:TTreview}
in an $\hbar$-independent form (i.e., letting $\hbar = 1$),
and show an $\hbar$-dependent formulation in the end of
this section.  Throughout this section, $s$ denotes a
discrete variable (``lattice coordinate'') with values
in $\bZ$, and $t = (t_1,t_2,\ldots)$ and
$\tbar = (\tbar_1,\tbar_2,\ldots)$ two sets of continuous
variables that play the role of ``time variables'' in the
Toda lattice hierarchy.

\subsection{Difference operators}

The Lax and Orlov-Shulman operators of the Toda lattice hierarchy
are difference operators of the form
\beqnarray
    L &=& e^{\rd_s} + \sum_{n=0}^\infty u_{n+1} e^{-n\rd_s},
                                                    \nonumber \\
    M &=& \sum_{n=1}^\infty n t_n L^n + s
        + \sum_{n=1}^\infty v_n L^{-n},
                                                    \nonumber \\
    \Lbar &=& \utilde_0 e^{\rd_s}
            + \sum_{n=0}^\infty \utilde_{n+1} e^{(n+2)\rd_s},
                                                    \nonumber \\
    \Mbar &=& - \sum_{n=1}^\infty n \tbar_n \Lbar^{-n} + s
            + \sum_{n=1}^\infty \vbar_n \Lbar^n,
\eeqnarray
where $e^{n\rd_s}$ are shift operators that act on a function
of $s$ as $ e^{n\rd_s} f(s) = f(s + n)$. The coefficients $u_n$,
$v_n$, $\utilde_n$ and $\vbar_n$ are functions of $(t,\tbar,q)$,
$u_n = u_n(t,\tbar,s)$, etc.
These operators obey the twisted canonical commutation relations
\beqn
    [L, M]= L, \quad [\Lbar, \Mbar]= \Lbar
                                   \label{eq:Toda-twistedCCR}
\eeqn
and the Lax equations
\beqnarray
    \frac{\rd L}{\rd t_n} = [B_n, L], &\quad&
    \frac{\rd L}{\rd \tbar_n} = [\Bbar_n, L],
                                              \nonumber \\
    \frac{\rd M}{\rd t_n} = [B_n, M], &\quad&
    \frac{\rd M}{\rd \tbar_n} = [\Bbar_n, M],
                                              \nonumber \\
    \frac{\rd \Lbar}{\rd t_n} = [B_n, \Lbar], &\quad&
    \frac{\rd \Lbar}{\rd \tbar_n} = [\Bbar_n, \Lbar],
                                              \nonumber \\
    \frac{\rd \Mbar}{\rd t_n} = [B_n, \Mbar], &\quad&
    \frac{\rd \Mbar}{\rd \tbar_n} = [\Bbar_n, \Mbar],
\eeqnarray
where the Zakharov-Shabat operators $B_n$ and $\Bbar_n$ are
given by
\beqn
    B_n = ( L^n )_{\ge 0}, \quad
    \Bbar_n = ( \Lbar^{-n} )_{< 0},
\eeqn
and $(\quad)_{\ge 0, <0}$ denotes the projection
\beqn
    ( \sum_n a_n e^{n \rd_s} )_{\ge 0} = \sum_{n \ge 0} a_n e^{n\rd_s},
    \quad
    ( \sum_n a_n e^{n \rd_s} )_{< 0} = \sum_{n < 0} a_n e^{n\rd_s}.
\eeqn
We call (\ref{eq:Toda-twistedCCR}) ``twisted'' because it is rather
the ``untwisted'' operators $ML^{-1}$ and $\Mbar\Lbar^{-1}$ that
give canonical conjugate variable of $L$ and $\Lbar$:
\beqn
    [L, ML^{-1}] = 1, \quad [\Lbar, \Mbar\Lbar^{-1}] = 1.
\eeqn

Another important set of difference operators are the so called
``dressing operators'' of the form
\beqnarray
    W = 1 + \sum_{n=1}^\infty w_n e^{-n\rd_s},
    &\quad&
    w_n = w_n(t,\tbar,s),
                                              \nonumber \\
    \Wbar = \wbar_0 + \sum_{n=1}^\infty \wbar_n e^{n\rd_s},
    &\quad&
    \wbar_n = \wbar_n(t,\tbar,s)
\eeqnarray
that ``undress'' the Lax and Orlov-Shulman operators as
\beqnarray
   L = W e^{\rd_s} W^{-1}, &\quad&
   M = W ( s + \sum_{n=1}^\infty n t_n e^{n\rd_s} ) W^{-1},
                                              \nonumber \\
   \Lbar = \Wbar e^{\rd_s} \Wbar^{-1}, &\quad&
   \Mbar = \Wbar ( s - \sum_{n=1}^\infty n \tbar_n e^{-n\rd_s} )
           \Wbar^{-1}.
                                   \label{eq:Toda-dressingRel}
\eeqnarray
This does not determine $W$ and $\Wbar$ uniquely, and one can
select a suitable pair of $W$ and $\Wbar$ such that the following
equations are satisfied:
\beqnarray
    \dfrac{\rd W}{\rd t_n}
      = - (W e^{n\rd_s} W^{-1})_{<0} W,        &&
    \dfrac{\rd W}{\rd \tbar_n}
      = (\Wbar e^{-n \rd_s} \Wbar^{-1})_{<0} W,
                                               \nonumber \\
    \dfrac{\rd \Wbar}{\rd t_n}
      = (W e^{n \rd_s} W^{-1})_{\ge 0} \Wbar,  &&
    \dfrac{\rd \Wbar}{\rd \tbar_n}
      = - (\Wbar e^{-n \rd_s} \Wbar^{-1})_{\ge 0} \Wbar.
                                       \label{eq:Toda-WWbarFlowEq}
\eeqnarray

These equations of flows in the space of dressing operators can
be ``linearized'' as follows.  A clue is the ``operator ratio''
\beqn
         U(t,\tbar) = W(t,\tbar)^{-1} \Wbar(t,\tbar)
                                      \label{eq:Toda-MulaseDecom}
\eeqn
of the dressing operators $W = W(t,\tbar)$ and $\Wbar = \Wbar(t,\tbar)$.
One can indeed easily see that $U = U(t,\tbar)$ satisfies the
``linear equations''
\beqn
    \dfrac{\rd U}{\rd t_n} = e^{n \rd_s} U, \quad
    \dfrac{\rd U}{\rd \tbar_n} = - U e^{-n \rd_s},
\eeqn
so that the flows in the space of $U$ operators are given by
simple exponential operators:
\beqn
    U(t,\tbar)
    = \exp\Bigl( \sum_{n=1}^\infty t_n e^{n \rd_s}\Bigr) U(0,0)
        \exp\Bigl( - \sum_{n=1}^\infty \tbar_n e^{-n \rd_s} \Bigr).
\eeqn
Furthermore, the passage from $(W,\Wbar)$ to $U$ is reversible.
Namely, given such an operator $U(t,\tbar)$, one can solve the
above ``factorization problem'' (\ref{eq:Toda-MulaseDecom})
to obtain two operators $W = W(t,\tbar)$ and $\Wbar = \Wbar(t,\tbar)$,
which then automatically satisfy (\ref{eq:Toda-WWbarFlowEq}).
This is a Toda lattice version of Mulase's factorization problem
for the KP hierarchy \cite{bib:Mulase}.  Actually, this factorization
problem is solved by reformulating the problem in the language of
infinite matrices.

\subsection{Infinite matrices}

Note the following one-to-one correspondence between the sets of
difference operators and of infinite ($\bZ \times \bZ$)
matrices:
\beqn
    \sum_n a_n(s) e^{n\rd_s}
    \longleftrightarrow
    \sum_n \diag[a_n(i)] \bLambda^n
\eeqn
where
\beqn
    \diag[a_n(i)] = \Bigl( a_n(i) \delta_{ij} \Bigr),
    \quad
    \bLambda^n = \Bigl( \delta_{i,j-n} \Bigr),
\eeqn
and the indices $i$ (row) and $j$ (column) run over $\bZ$. (This
$\bLambda$ should not be confused with the finite matrix $\Lambda$
in the partition function of generalized Kontsevich models!)
The projectors $(\quad)_{\ge 0,<0}$ are replaced by the projectors
onto upper and (strictly) lower triangular matrices:
\beqn
    ( A )_{\ge 0} = \Bigl( \theta(j-i) a_{ij} \Bigr), \quad
    ( A )_{< 0} = \Bigl( (1 - \theta(j-i)) a_{ij} \Bigr).
\eeqn
All the Lax, Orlov-Shulman, Zakharov-Shabat and dressing operators
have their counterparts in infinite matrices,
\beqn
    L     \leftrightarrow \bL,    \quad
    M     \leftrightarrow \bM,    \quad
    \Lbar \leftrightarrow \bLbar, \quad
    \Mbar \leftrightarrow \bMbar, \quad
    W     \leftrightarrow  \bW,    \quad
    \Wbar \leftrightarrow \bWbar,
\eeqn
and they obey the same equations as their counterparts.
For instance, dressing relation (\ref{eq:Toda-dressingRel})
turns into a matrix relation of the form
\beqnarray
   \bL = \bW \bLambda \bW^{-1}, &\quad&
   \bM = \bW ( \bDelta + \sum_{n=1}^\infty
                          n t_n \bLambda^n ) \bW^{-1},
                                              \nonumber \\
   \bLbar = \bWbar \bLambda \bWbar^{-1}, &\quad&
   \bMbar = \bWbar ( \bDelta - \sum_{n=1}^\infty n
                          \tbar_n \bLambda^{-n} ) \bWbar^{-1}.
\eeqnarray
Here $\bDelta$ is the infinite matrix
\beqn
    \bDelta = \Bigl( i \delta_{ij} \Bigr)
\eeqn
that represents the multiplication operator $s$.
Similarly, the difference operator $U(t,\tbar)$ has an infinite
matrix counterpart $\bU(t,\tbar)$, and its time evolutions are
generated by exponential matrices:
\beqn
    \bU(t,\tbar) = \exp( \sum_{n=1}^\infty t_n \bLambda^n ) \bU(0,0)
                   \exp( -\sum_{n=1}^\infty \tbar_n \bLambda^{-n} ).
\eeqn

Factorization relation (\ref{eq:Toda-MulaseDecom}) is now converted
into a factorization problem of infinite matrices of the form
\beqn
    \bU(t,\tbar) = \bW(t,\tbar)^{-1} \bWbar(t,\tbar).
                                       \label{eq:Toda-GaussDecom}
\eeqn
This is an infinite matrix
version of the Gauss decomposition, because $\bW = \bW(t,\tbar)$
and $\bWbar = \bWbar(t,\tbar)$ are lower and upper triangular
matrices. As in the ordinary finite dimensional cases, this Gauss
decomposition can be solved explicitly by (an infinite matrix
version of) the Cramer formula \cite{bib:TodaIVP}. In particular,
the matrix elements $w_n(t,\tbar,s)$ and $\wbar_n(t,\tbar,s)$ can
be written as a quotient of two semi-infinite determinants. The
denominators eventually turn out to give $\tau$ functions of the
Toda lattice hierarchy:
\beqn
    \tau(t,\tbar,s) = \det\Bigl( u_{ij}(t,\tbar) \
                                 (-\infty <i,j<s) \Bigr).
                                      \label{eq:Toda-TauDet}
\eeqn
This gives a Toda lattice version of a similar formula in the
KP hierarchy \cite{bib:KP}. The matrix elements $u_{ij}(t,\tbar)$
of $\bU(t,\tbar)$ are connected with their ``initial values''
$u_{ij}(0,0)$ by
\beqn
    u_{ij}(t,\tbar) = \sum_{m,n=0}^\infty
                        S_m(t) u_{i+m,j+n}(0,0) S_n(-\tbar) ,
\eeqn
where $S_n$ are the fundamental Schur functions
\beqn
    \sum_{n=0}^\infty S_n(t) \lambda^n =
     \exp\Bigl( \sum_{n=1}^\infty t_n \lambda^n \Bigr).
\eeqn
Eventually, the infinite matrix (a $GL(\infty)$ element) $\bU(0,0)$
persists as arbitrary constants in a general solution of the
Toda lattice hierarchy.  It is thus the $GL(\infty)$ group
that plays the role of the Sato Grassmannian in the Toda lattice
hierarchy.

\subsection{Constraints}

``String equations'' in our concern are derived from linear
relations of the form
\beqn
    A(\bDelta,\bLambda) \bU(0,0) = \bU(0,0) \Abar(\bDelta,\bLambda),
\eeqn
among the matrix elements of $\bU(0,0)$. Here $A(\bDelta,\bLambda)$
and $\Abar(\bDelta,\bLambda)$ are linear combinations of monomials
of $\bDelta$, $\bLambda$ and $\bLambda^{-1}$:
\beqnarray
    A(\bDelta,\bLambda)
    &=& \sum_{m,n} a_{mn} \bDelta^m \bLambda^n,
                                                   \nonumber \\
    \Abar(\bDelta,\bLambda)
    &=& \sum_{m,n} \abar_{mn} \bDelta^m \bLambda^n,
\eeqnarray
$m$ runing over nonnegative integers and $n$ over all
integers.  A fundamental result is the following
\cite{bib:TTToda,bib:ASvM,bib:NTTceq1}:

\begin{theorem} The above linear relations of $\bU(0,0)$ is
equivalent to the constraint
\beqn
    A(\bM,\bL) = \Abar(\bMbar,\bLbar)
                                  \label{eq:Toda-LaxConstraint}
\eeqn
of the Lax and Orlov-Shulman operators, where
\beqn
    A(\bM,\bL)
      = \sum_{m,n} a_{mn} \bM^m \bL^n,  \quad
    \Abar(\bMbar,\bLbar)
      = \sum_{m,n} \abar_{mn} \bMbar^m \bLbar^n.
\eeqn
\end{theorem}

These Constraints may be interpreted as a fixed point condition
under $W_{1+\infty}$ symmetries of the Toda lattice hierarchy
\cite{bib:TTToda,bib:ASvM,bib:NTTceq1}. These matrices are
in one-to-one correspondence with difference operators,
\beqn
    A(\bDelta,\bLambda) \leftrightarrow A(s,e^{\rd_s}), \quad
    \Abar(\bDelta,\bLambda) \leftrightarrow \Abar(s,e^{\rd_s}),
\eeqn
and give a closed Lie algebra with the fundamental commutation
relation
\beqn
    [\bDelta, \bLambda] = \bDelta \ \leftrightarrow \
    [e^{\rd_s}, s] = e^{\rd_s}.
\eeqn
This is essentially a (centerless) $W_{1+\infty}$ algebra
with generators
\beqn
    W^{(k)}_n = (\bDelta\bLambda^{-1})^k \bLambda^{n+k}
    \ \leftrightarrow \ (s e^{-\rd_s})^k e^{(n+k)\rd_s}.
\eeqn
It is this $W_{1+\infty}$ algebra that underlies the above
constraints.

These constraints can be further converted into linear
constraints for the $\tau$ functions of the form
\cite{bib:TTToda,bib:ASvM,bib:NTTceq1}
\beqn
    X_A \tau(t,\tbar,s)
    = \Xbar_\Abar \tau(t,\tbar,s) + \const \tau(t,\tbar,s),
                                   \label{eq:Toda-TauConstraint}
\eeqn
where $X_A = X_A(t,s,\rd_t)$ and $\Xbar_\Abar(\tbar,s,\rd_\tbar)$
are linear differential operators in $t$ and $\tbar$ that represent
$W_{1+\infty}$ symmetries acting on the $\tau$ functions, and
``$\const$'' is a constant that also depends on $A$ and $\Abar$.
This constant may emerge due to a nonvanishing central charge
in the $W_{1+\infty}$ algebra of symmetries acting on the
$\tau$ functions.  Fixing this constant requires a subtle
trick \cite{bib:FKN,bib:vandeLeur}. We shall not deal with
this issue in this paper.

\subsection{$\tau$ functions in Miwa variables}

A clue connecting generalized Kontsevich models and the KP
hierarchy \cite{bib:Kontsevich,bib:AvM} is the Miwa variable
representation of the KP $\tau$ function. If the matrix
$\bU(0,0)$ is upper or lower triangular, a similar
representation of the Toda lattice $\tau$ functions
\cite{bib:GKMvsToda} is also available as we show below.
(In fact, such a representation persists without this
condition, though the relation to $\bU$ then becomes more
complicated.)

\begin{theorem} (i) Suppose that $\bU(0,0)$ is lower triangular.
Then by the Miwa transformation
\beqn
    t_n = \frac{1}{n} \sum_{i=1}^N \lambda_i^{-n}
\eeqn
of $t$ with arbitrary parameters $\lambda_i$, the $\tau$
functions $\tau(t,\tbar,s)$ can be written
\beqn
    \tau(t,\tbar,s)
  = \dfrac{ \det\Bigl( u_{s-j}(0,\tbar,\lambda_i) \
                          (1 \le i,j \le N) \Bigr) }
            { \det\Bigl( \lambda_i^{-(s-j)-1} \
                          (1 \le i,j \le N) \Bigr) }
    \prod_{i=-\infty}^{s-N-1} u_{ii},
                                    \label{eq:Toda-TauMiwaU}
\eeqn
where $u_j(0,\tbar,\lambda)$ are the following generating
functions of matrix elements $u_{ij}(0,\tbar)$ of $\bU(0,\tbar)$:
\beqn
    u_j(0,\tbar,\lambda)
    =  \sum_{i=j}^\infty \lambda^{-i-1} u_{ij}(0,\tbar),
\eeqn
\newline
(ii) Suppose that $\bU(0,0)$ is upper triangular. Then by
the Miwa transformation
\beqn
    \tbar_n = - \frac{1}{n} \sum_{i = 1}^N \mu_i^n
\eeqn
of $\tbar$ with arbitrary parameters $\mu$, the $\tau$
function can be written
\beqn
    \tau(t,\tbar,s)
  =
    \dfrac{ \det\Bigl( \ubar_{s-i}(t,0,\mu_j) \
                       (1 \le i,j \le N) \Bigr) }
          { \det\Bigl( \mu_j^{s-i} \
                       (1 \le i,j \le N) \Bigr) }
    \prod_{i=-\infty}^{s-N-1} u_{ii},
                                \label{eq:Toda-TauMiwaUbar}
\eeqn
where $\ubar_i(t,0,\mu)$ are the following generating functions
of the matrix elements $u_{ij}(t,0)$ of $\bU(t,0)$:
\beqn
    \ubar_i(t,0,\mu) = \sum_{j=i}^\infty u_{ij}(t,0) \mu^j.
\eeqn
\end{theorem}

The infinite products of $u_{ii}$ and $\ubar_{ii}$ in the above
formulae are interpreted as follows. Note that the matrices
$\bU(0,\tbar)$ in (i) and $\bU(t,0)$ in (ii) are still triangular.
The Laurent series $u_j(0,\tbar,\lambda)$ and $\ubar_(t,0,\,u)$
thereby take such a form as
\beqn
    u_j(0,\tbar,\lambda) =
        \sum_{i=j}^\infty \lambda^{-i-1} u_{ij}(0,\tbar),
    \qquad
    \ubar_i(t,0,\mu) = \sum_{j=i}^\infty u_{ij}(t,0) \mu^j.
\eeqn
As opposed to an ordinary setting \cite{bib:GKMvsToda},
we do not assume that the leading coefficients are normalized
to be $1$.  The infinite product then arises.  This factor is
essentially the same as encountered in the treatment of
semi-infinite determinant (\ref{eq:Toda-TauDet})
 \cite{bib:TodaIVP}, and can be interpreted as:
\beqn
    \prod_{-\infty}^n u_{ii}
    = \left\{ \begin{array}{ll}
           \const u_{00}\ldots u_{nn}              &  (n \ge 0) \\
           \const                                  &  (n = -1)  \\
           \const /(u_{n+1,n+1} \ldots u_{-1,-1}) &  (n \le -2)
     \end{array} \right.
\eeqn
with an overall renormalization constant ``$\const$''

\subsection{$\hbar$-dependent formulation}

An $\hbar$-dependent formulation of Toda lattice hierarchy can
be achieved by inserting $\hbar$ in front of all derivatives in
the previous equations as:
\beqn
    \dfrac{\rd}{\rd t_n}     \to \dfrac{\rd}{\rd t_n}, \quad
    \dfrac{\rd}{\rd \tbar_n} \to \dfrac{\rd}{\rd \tbar_n}, \quad
    e^{\rd_s}                \to e^{\hbar\rd_s}.
\eeqn
The discrete variable $s$ now takes values in $\hbar \bZ$.
Accordingly, Lax and Orlov-Shulman operators are difference
operators of the form
\beqnarray
    L &=& e^{\hbar\rd_s} + \sum_{n=0}^\infty u_{n+1} e^{-n\hbar\rd_s},
                                                    \nonumber \\
    M &=& \sum_{n=1}^\infty n t_n L^n + s
        + \sum_{n=1}^\infty v_n L^{-n},
                                                    \nonumber \\
    \Lbar &=& \utilde_0 e^{\hbar\rd_s}
            + \sum_{n=0}^\infty \utilde_{n+1} e^{(n+2)\hbar\rd_s},
                                                    \nonumber \\
    \Mbar &=& - \sum_{n=1}^\infty n \tbar_n \Lbar^{-n} + s
            + \sum_{n=1}^\infty \vbar_n \Lbar^n,
\eeqnarray
where the coefficients are functions of $(\hbar,t,\tbar,s)$,
and obey the twisted canonical commutation relations
\beqn
   [ L , M ] = \hbar L, \quad [ \Lbar , \Mbar ] = \hbar \Lbar,
\eeqn
and Lax equations
\beqnarray
    \hbar \frac{\rd L}{\rd t_n} = [B_n, L], &\quad&
    \hbar \frac{\rd L}{\rd \tbar_n} = [\Bbar_n, L],
                                              \nonumber \\
    \hbar \frac{\rd M}{\rd t_n} = [B_n, M], &\quad&
    \hbar \frac{\rd M}{\rd \tbar_n} = [\Bbar_n, M],
                                              \nonumber \\
    \hbar \frac{\rd \Lbar}{\rd t_n} = [B_n, \Lbar], &\quad&
    \hbar \frac{\rd \Lbar}{\rd \tbar_n} = [\Bbar_n, \Lbar],
                                              \nonumber \\
    \hbar \frac{\rd \Mbar}{\rd t_n} = [B_n, \Mbar], &\quad&
    \hbar \frac{\rd \Mbar}{\rd \tbar_n} = [\Bbar_n, \Mbar].
\eeqnarray
$B_n$ and $\Bbar_n$ are defined in the the same way as in the
case of $\hbar = 1$.

Note that such an $\hbar$-dependent Toda lattice hierarchy emerges
if one starts from an $\hbar$-independent formulation and rescales
variables as
\beqn
    t_n     \to \hbar^{-1} t_n, \quad
    \tbar_n \to \hbar^{-1} \tbar_n, \quad
    s       \to \hbar^{-1} s
\eeqn
and
\beqn
    M     \to \hbar^{-1} M, \quad
    \Mbar \to \hbar^{-1} \Mbar.
\eeqn
This is indeed the case for solutions that we shall construct
in subsequent sections. The $\hbar$-dependent formulation, however,
also admits solutions that cannot be obtained by the above rescaling
of $(t,\tbar,s)$.

\section{Generalized Kontsevich models}

Our first example is the generalized Kontsevich models that
have been studied in the context of the KP hierarchy.
We introduce ``negative times'' $\tbar$ and reconsider
these models in the language of the Toda lattice hierarchy.
The idea is more or less parallel to the ITEP-Lebedev group
\cite{bib:GKMvsToda}, who however treated these models as
solutions of a ``forced'' hierarchy on a semi-infinite lattice
$\bZ_{\ge 0}$. We rather attempt to interpret these models
as solutions of the hierarchy on the bi-infinite lattice $\bZ$.

\subsection{Partition function as KP $\tau$ function}

In an $\hbar$-dependent formulation, the partition function of
the $p$-th generalized Kontsevich model is given by
\beqn
    Z(\Lambda) = C(\Lambda) \int dM \exp \hbar^{-1} \Tr \Bigl(
                     \Lambda^p M - \frac{1}{p+1} M^{p+1}\Bigr),
\eeqn
where the normalization constant $C(\Lambda)$ is given by
\beqnarray
    C(\Lambda)
    &=& \const \prod_{i>j} \dfrac{ \lambda_i^p - \lambda_j^p }
                                 { \lambda_i - \lambda_j }
                                                    \nonumber \\
    && \times
       (\det \Lambda)^{(p-1)/2} \exp \hbar^{-1} \Tr \Bigl(
        \frac{p}{p+1} \Lambda^{p+1} \Bigr)
\eeqnarray
By the standard method using the Harish-Chandra-Itzykson-Zuber
formula \cite{bib:Morozov}, the partion function $Z(\Lambda)$
can be rewritten as a quotient of two determinants,
\beqn
    Z(\Lambda)
    = \dfrac{ \det\Bigl( u_{-j}(\lambda_i) \
              (1 \le i,j \le N) \Bigr) }
            { \det\Bigl( \lambda_i^{j-1} \
              (1 \le i,j \le N) \Bigr) },
                                   \label{eq:GKMZasKPtau}
\eeqn
where
\beqnarray
    u_j(\lambda)
      &=& c(\lambda) \int d\mu \mu^{-j-1} \exp \hbar^{-1} \Bigl(
                  \lambda^p \mu - \frac{1}{p+1} \mu^{p+1} \Bigr),
                                                    \nonumber \\
    c(\lambda)
      &=& \const \mu^{(p-1)/2} \exp \hbar^{-1} \Bigl(
                   \frac{p}{p+1} \lambda^{p+1} \Bigr).
\eeqnarray
The right hand side of the definition of $u_j(\lambda)$ is an
integral over the whole real axis.  By the standard saddle point
method, one can show that $u_j(\lambda)$ has symptotic expansion of
the form
\beqn
    u_j(\lambda) \sim \sum_{i=j}^\infty \lambda^{-i-1} u_{ij}
\eeqn
as $\lambda \to + \infty$. It is these asymptotic series
rather than the functions $u_j(\lambda)$ themselves that are
eventually relevant for the interpretation of $Z(\Lambda)$ as
a KP $\tau$ function.  Namely, we insert this asymptotic
expansion into the right hand side of (\ref{eq:GKMZasKPtau})
and consider it as a function (or, rather, formal power series)
of $t$ by the Miwa transformation
\beqn
    t_n = \frac{\hbar}{n} \sum_{i=1}^N \lambda_i^{-n}
        = \frac{\hbar}{n} \Tr \Lambda^{-n}.
\eeqn

\subsection{Extension to Toda lattice $\tau$ function}

We now extend the above construction to $u_j$'s with negative
index $j$. Apparently, this appears problematical, because
the integral (over the real axis) becomes singular at $\mu = 0$;
actually, what we need is asymptotic series rather the functions
$u_j(\lambda)$ themselves. As we shall discuss later in a more
general setting, the asymptotic expansion as $\lambda \to + \infty$
is determined only by a small neighborhood of the saddle point at
$\mu = \lambda$. The other part of the path of integration may be
slightly deformed to avoid the singularity at $\mu = 0$,
or even cut off!  With this deformation or cut-off,
the integral itself will be varied, but the difference is
subdominant and does not affect the asymptotic series. Thus
upon suitably modifying the definition of $u_j(\lambda)$, we can
show that $u_j(\lambda)$ for $j < 0$, too, has asymptotic
expansion of the form
\beqn
    u_j(\lambda) \sim \sum_{i=j}^\infty \lambda^{-i-1} u_{ij}.
\eeqn
We can thus define $u_{ij}$ for all integers, and hence a
$\bZ \times \bZ$ matrix $\bU = \bU(0,0)$.
According to the general constrution in the last section,
this determines a Toda lattice $\tau$ function $\tau(t,\tbar,s)$.

Since $\bU(0,0)$ is lower triangular, the $\tau$ function should
have a finite determinant representation by the Miwa transformation
\beqn
    t_n = \frac{\hbar}{n} \sum_{i=1}^N \lambda_i^{-n}
        = \frac{\hbar}{n} \Tr \Lambda^{-n}.
\eeqn
One can easily see that the generating functions $u_j(0,\tbar,\lambda)$
is given by
\beqn
    u_j(0,\tbar,\lambda)
      =  c(\lambda) \int d\mu \mu^{-j-1} \exp \hbar^{-1} \Bigl(
                  \lambda^p \mu - \frac{1}{p+1} \mu^{p+1}
         - \sum_{n=1}^\infty \tbar_n \mu^{-n} \Bigr)
\eeqn
with the same normalization factor $c(\lambda)$ and the same path
of integration (deformed or cut off in the aforementioned sense)
as in the definition of $u_j(\lambda) = u_j(0,0,\lambda)$.
The $\tau(t,\tbar,s)$ can eventually be written
\beqn
    \tau(t,\tbar,s)
    = \dfrac{ \det\Bigl( u_{\hbar^{-1}s-j}(0,\tbar,\lambda_i) \
              (1 \le i,j \le N) \Bigr) }
            { \det\Bigl( \lambda_i^{\hbar^{-1}s - j -1} \
              (1 \le i,j \le N) \Bigr) }.
                                     \label{eq:KGMZasTodatau}
\eeqn
(Note that, along with $t$ and $\tbar$, the lattice coordinate
$s$, too, has to be rescaled as $s \to \hbar^{-1}s$ in an
$\hbar$-dependent formulation of the Toda lattice hierarchy.
$s$ now runs over the $\hbar$-spaced lattice $\hbar \bZ$.)

Formally, this $\tau$ function can be derived from the following
matrix integral that extends $Z(\Lambda)$:
\beqnarray
    Z(\Lambda,\tbar,s)
    &=& C(\Lambda) \int dM \exp \hbar^{-1} \Tr \Bigl(
                  \Lambda^p M - \frac{1}{p+1} M^{p+1}
                                           \nonumber \\
    &&  - \sum_{n=1}^\infty \tbar_n M^{-n} + s \log M\Bigr).
\eeqnarray

\subsection{String equations}

It is now rather straightforward to derive string equations ---
we first derive linear relations among $u_{ij} = u_{ij}(0,0)$,
then convert them into relations among $(L,M,\Lbar,\Mbar)$.

Linear relations can be derived from linear functional or
differential relations among $u_j$'s.  This is just to apply
well known calculations \cite{bib:GKMvsToda,bib:Morozov}
to $u_j$'s with negative indices as follows.

First, by integrating by part,
\beqnarray
    &&  \lambda^p u_j(\lambda)
                                         \nonumber \\
    &=& c(\lambda) \int d\mu \hbar \frac{\rd}{\rd \lambda}
        \Bigl(\exp( \hbar^{-1} \lambda^p \mu ) \Bigr)
        \times \mu^{-j-1} \exp \hbar^{-1}
        \Bigl( - \frac{1}{p+1} \mu^{p+1} \Bigr)
                                         \nonumber \\
    &=& c(\lambda) \int d\mu \Bigl( \hbar (j+1) \mu^{-j-2}
        + \mu^{-j+p-1} \Bigr)
        \times \exp \hbar^{-1} \Bigl( \lambda^p \mu
               - \frac{1}{p+1} \mu^{p+1} \Bigr)
                                         \nonumber \\
    &=& \hbar (j+1) u_{j+1}(\lambda) + u_{j-p}(\lambda).
\eeqnarray
Here we have implicitly assumed that the path of integration
connects two points at infinity, so that no boundary term
emerges.  If the path has a finite endpoint, such as the
semi-infinite line $[\epsilon,\infty)$, $\epsilon > 0$, the
boundary term is subdominant compared to the main asymptotic
series arising from the saddle point at $\mu = \lambda$.
Since we shall eventually derive linear relations among
$u_{ij}$'s, such subdominant terms are negligible.

Similarly (but without integration by part),
\beqnarray
    &&  u_{j-1}(\lambda)
                                        \nonumber \\
    &=& c(\lambda) \int d\mu \hbar \lambda^{1-p} \frac{1}{p}
        \frac{\rd}{\rd \lambda} \Biggl( \mu^{-j-1}
        \exp \hbar^{-1} \bigl( \lambda^p \mu
              - \frac{1}{p+1} \mu^{p+1} \Bigr) \Biggr)
                                        \nonumber \\
    &=& c(\lambda) \hbar \lambda^{1-p} \frac{1}{p}
        \dfrac{\rd}{\rd \lambda} \Bigl( c(\lambda)^{-1}
        u_j(\lambda) \Bigr)
                                        \nonumber \\
    &=& \Bigl( \hbar \lambda^{1-p} \frac{1}{p}
                  \frac{\rd}{\rd \lambda}
        - \hbar \frac{p-1}{2p} \lambda^{-p} + \lambda
        \Bigr) u_j(\lambda).
\eeqnarray

These relations imply the linear relations
\beqnarray
    u_{i+p,j} &=& \hbar (j+1) u_{i,j+p} + u_{i,j-1},
                                         \nonumber \\
    u_{i,j-1} &=& \hbar( - \frac{i}{p} + \frac{p-1}{2p} )u_{i-p,j}
                  + u_{i+1,j}
\eeqnarray
among the coefficients of asymptotic expansion. In terms of
$\bU = \bU(0,0)$, they can be rewritten
\beqnarray
    \bLambda^p \bU &=& \bU ( \hbar \bDelta \bLambda^{-1} + \bLambda^p),
                                          \nonumber \\
    \bU \bLambda &=& ( - \frac{\hbar}{p} \bDelta \bLambda^{-p}
                      + \hbar \frac{p-1}{2p} \bLambda^{-p}
                      + \bLambda) \bU.
\eeqnarray

Finally, we replace
\beqn
    \bLambda      \longrightarrow  L, \Lbar, \quad
    \hbar \bDelta \longrightarrow  M, \Mbar
\eeqn
(taking into account the rescaling due to the presence of
$\hbar$) and obtain the following result on string equations:

\begin{theorem} The Lax and Orlov-Shulman operators of
this solution obey the string equations
\beqnarray
    L^p &=& \Mbar \Lbar^{-1} + \Lbar^p,
                                        \nonumber \\
    \Lbar &=& - \frac{1}{p} M L^{-p}
              + \hbar \frac{p-1}{2p} L^{-p} + L.
\eeqnarray
\end{theorem}

\subsection{Generalized Kontsevich model at $p = -1$}

Let us consider the case of $p = -1$.  The string equations
then become
\beqnarray
    L^{-1} &=& \Mbar \Lbar^{-1} + \Lbar^{-1}, \nonumber \\
    \Lbar  &=& M L + \hbar L  + L.
                                 \label{eq:StringEq(-1,1)}
\eeqnarray
As we shall argue in the next section, these string equations
do not agree with string equations of two-dimensional
(or $c = 1$) strings (compactified at self-dual radius).

This is a puzzling consequence of our analysis. Some of
recent studies on the two-dimensional topological string
theory \cite{bib:Ozetal} are based on the hypothesis that
the topological $A_k$ model at $k = -3$ (which corresponds
to $p = -1$) gives the two-dimensional string theory.
Our result implies that such an identification is somewhat
problematical.

The model at $p = -1$, as Dijkgraaf et al. \cite{bib:DMP}
remarked, is rather related to the Penner model.
The partition function for $p = -1$ is given by
\beqn
    Z(\Lambda,\tbar,0)
      = C(\Lambda) \int dM \exp \hbar^{-1} \Tr \Bigl(
                  \Lambda^{-1} M - \log M
        - \sum_{n=1}^\infty \tbar_n M^{-n} \Bigr).
                                 \label{eq:GKM(-1)}
\eeqn
If $\tbar = 0$,
this function factorizes into a power of $\det\Lambda$ and
$Z(1,0,0)$, and that the second factor $Z(1,0,0)$ coincides
with the partition function of the Penner model \cite{bib:Penner}.
Curiously, however, Dijkgraaf et al. identified this matrix
integral with the $c = 1$ string partition function. Actually,
as we have mentioned above, this matrix integral does not
correspond to the $c = 1$ string partition function.

What physical meaning do the ``negative times'' $\tbar$ possess?
It seems likely that $\tbar$ are nothing else but the coupling
constants of ``anti-states'' that Montano and Rivlis
\cite{bib:MontanoRivlis} postulate in their topological
interpretation of Ward identities for $(1,q)$ models. This is
in fact a main idea that lies in the heart of the work of
the aforementioned studies \cite{bib:Ozetal} on the
two-dimensional topological string theory. Although we
consider their interpretation of the $(1,q=-1)$ model rather
problematical, their work contains many intriguing ideas.
We shall return to this point in the final section.

\section{Compactified $c = 1$ string theory}

Our second example is the compactified $c = 1$ string theory
formulated in the language of the Toda lattice hierarchy
\cite{bib:DMP,bib:Marshakov,bib:Nakatsu}. This string theory
possesses a discrete parameter $\beta = 1,2,\ldots$, and
$\beta = 1$ corresponds to the case at self-dual radius.

\subsection{Partition function of $c = 1$ strings}

The partition function of compactified $c = 1$ strings becomes
the Toda lattice $\tau$ function with a diagonal $\bU = \bU(0,0)$
matrix of the form \cite{bib:DMP,bib:Marshakov,bib:Nakatsu}
\beqn
    u_{ij}
    = (- \hbar)^{ (i + \frac{1}{2})/\beta}
       \frac{ \Gamma\Bigl(\frac{1}{2} - \frac{1}{\hbar}
               + \frac{i + \frac{1}{2}}{\beta} \Bigr) }
             { \Gamma\Bigl(\frac{1}{2} - \frac{1}{\hbar} \Bigr) }
       \delta_{ij}.
\eeqn
In fact, this is a somewhat simplified version of a true $c = 1$
string partition function \cite{bib:DMP}.  The above definition
is substantially the same as used by Nakatsu \cite{bib:Nakatsu}
and leads to the same string equations (except that $\hbar$
in his definition is replaced by $-\hbar$ here). String equations
of the original $c = 1$ string partition function are presented
in Ref. \cite{bib:NTTceq1} in a more general setting including
black hole backgrounds.

Since $\bU = \bU(0,0)$ is a diagonal matrix, the $\tau$ functions
have two Miwa variable representations with respect to $t$ and $\tbar$.
In both representations, the $\tau$ functions are written in terms
of a finite determinant including the generating functions
$u_j(0,\tbar,\lambda)$ and $\ubar_i(t,0,\mu)$. By the familiar
integral representation of the Gamma function, one can easily
obtain the following integral representation of these generating
functions:
\beqnarray
    u_j(0,\tbar,\lambda)
    &=& \frac{ \beta (- \hbar)^{\frac{1}{\hbar} - \frac{1}{2}} }
              { \Gamma\Bigl( \frac{1}{2} - \frac{1}{\hbar} \Bigr) }
        \lambda^{(\beta-1)/2} \exp(- \hbar^{-1} \beta \log\lambda)
                                                 \nonumber \\
    &&  \times \int_0^\infty d\mu \mu^{(-\beta-1)/2 - j - 1}
        \exp \hbar^{-1} \Bigl( (\lambda/\mu)^\beta
         + \beta \log \mu - \sum_{n=1}^\infty \tbar_n \mu^{-n} \Bigr),
                                                 \nonumber \\
    \ubar_i(t,0,\mu)
    &=& \frac{ \beta (- \hbar)^{\frac{1}{\hbar} - \frac{1}{2}} }
              { \Gamma\Bigl( \frac{1}{2} - \frac{1}{\hbar} \Bigr) }
        \mu^{(-\beta-1)/2} \exp( \hbar^{-1} \beta \log \mu )
                                                 \nonumber \\
    &&  \times \int_0^\infty d\lambda \lambda^{(\beta-1)/2 + i}
              \exp \hbar^{-1} \Bigl( (\lambda/\mu)^\beta
        - \beta \log \lambda + \sum_{n=1}^\infty t_n \lambda^n \Bigr).
\eeqnarray

\subsection{Integral representation of Kontsevich type}

If $\beta = 1$ (i.e., in the case at self-dual radius),
the $\tau$ function has a matrix integral representation
of Kontsevich type.  For simplicity, let us consider the
case of $s = \hbar (N-1)$ (though a similar matrix integral
representation persists in a general case).  By the Miwa
transformation
\beqn
    \tbar_n = - \frac{\hbar}{n} \sum_{i = 1}^N \mu_i^n
            = - \frac{\hbar}{n} \Tr M^n,
\eeqn
the $\tau$ function $\tau(t,\tbar,\hbar (N-1))$ can be written
\beqn
    \tau(t, \tbar, \hbar (N-1))
    = \const
      \dfrac{ \det\Bigl( \ubar_{N-i-1}(t,0,\mu_j) \
                         (1 \le i,j \le N) \Bigr) }
            { \det\Bigl( \mu_j^{N-i-1} \
                         (1 \le i,j \le N) \Bigr) }.
\eeqn
As in the case of generalized Kontsevich models, this can be
converted into a matrix integral of the form
\beqn
    \tau(t,\tbar,\hbar (N-1))
    = \Cbar(M) \int d \Lambda \exp \hbar^{-1} \Tr \Bigl(
        \Lambda M^{-1} - \log \Lambda
        + \sum_{n=1}^\infty t_n \Lambda^n \Bigr)
\eeqn
with a suitable normalization factor $\Cbar(M)$.

Although very similar, this matrix integral is substantially
different from the generalized Kontsevich model at $p = -1$.
The latter rather corresponds to $\beta = -1$. (Mathematically,
the present construction is also valid for $\beta = -1,-2,\ldots$.)
If $\beta = -1$, one can indeed derive a matrix integral
representation of the form
\beqn
    \tau(t,\tbar,0)
    = C(\Lambda) \int dM \exp \hbar^{-1} \Tr \Bigl(
        \Lambda^{-1} M - \log M
        - \sum_{n=1}^\infty \tbar_n M^{-n} \Bigr)
\eeqn
in the Miwa variables
\beqn
    t_n = \frac{\hbar}{n} \sum_{i=1}^N \lambda_i^{-n}
        = \frac{\hbar}{n} \Tr \Lambda^{-n}.
\eeqn
(We have put $s = 0$ just for simplicity. A similar expression
persists for a general value of $s$.)  This is nothing else
but the generalized Kontsevich model at $p = -1$ given by
(\ref{eq:GKM(-1)}).

Thus, as opposed to the claim of Dijkgraaf et al. \cite{bib:DMP},
we conclude that the generalized Kontsevich model at $p = -1$
does not correspond to the $c = 1$ string theory of $\beta = 1$,
but falls into the case of the strange value $\beta = -1$.
We shall reconfirm this puzzling conclusion in the language of
string equations below.

\subsection{String equations}

By the recursion relation $\Gamma(x+1) = x \Gamma(x)$ of the Gamma
function, one can easily derive the following algebraic relations
for $\bU = \bU(0,0)$:
\beqnarray
    \bLambda^\beta \bU &=& \bU \Bigl( - \frac{\hbar}{\beta} \bDelta
      - \hbar \frac{\beta+1}{2\beta} + 1 \Bigr) \bLambda^\beta,
                                         \nonumber \\
    \bU \bLambda^{-\beta} &=& \Bigl( - \frac{\hbar}{\beta} \bDelta
      + \hbar \frac{\beta-1}{2\beta} + 1 \Bigr) \bLambda^{-\beta} \bU.
\eeqnarray
Here $\bLambda$ and $\bDelta$ are the infinite matrices introduced
in Section 2. To obtain the associated string equations, we have
only to resort to the substitution rule
\beqn
    \bLambda \to L, \Lbar, \quad
    \hbar \bDelta \to M, \Mbar.
\eeqn
We can thus reproduce the following result of Nakatsu
\cite{bib:Nakatsu} (who derives this result by a somewhat
different method).

\begin{theorem}
The Lax and Orlov-Shulman operators of this solution obey
the string equations
\beqnarray
    L^\beta &=& \Bigl( - \frac{1}{\beta} \Mbar
      - \hbar \frac{\beta+1}{2\beta} + 1 \Bigr) \Lbar^\beta,
                                          \nonumber \\
    \Lbar^{-\beta} &=& \Bigl( - \frac{1}{\beta} M
      + \hbar \frac{\beta-1}{2\beta} + 1) L^{-\beta}.
\eeqnarray
\end{theorem}

As promised, this result clearly shows that string equations
(\ref{eq:StringEq(-1,1)}) of the generalized Kontsevich model
at $p = - 1$ agree with none of the $c = 1$ string theory with
$\beta \ge 1$. The only possible value is $\beta = -1$.

One can rewrite these string equations into linear constraints
($W_{1+\infty}$ constraints) on the $\tau$ functions.
To do this, just take the $n$-th power ($n = 1,2,\ldots)$
of both hand sides of the above string equations as
\beqn
    L^{n\beta} = (\cdots)^n, \quad
    \Lbar^{n\beta} = (\cdots)^n,
\eeqn
and apply the relation between nonlinear constraints
(\ref{eq:Toda-LaxConstraint}) on $(L,M,\Lbar,\Mbar)$
and linear constraints (\ref{eq:Toda-TauConstraint})
on the $\tau$ function. If $\beta = 1,2,\ldots$,
one will thus obtain a set of $W_{1+\infty}$
constraints of the form
\beqnarray
    \dfrac{\rd \tau(t,\tbar,s)}{\rd t_{n\beta}}
    &=& \Xbar_{\beta,n}(\tbar,s,\rd_\tbar) \tau(t,\tbar,s),
                                       \nonumber \\
    \dfrac{\rd \tau(t,\tbar,s)}{\rd \tbar_{n\beta}}
    &=& X_{\beta,n}(t,s,\rd_t) \tau(t,\tbar,s),
\eeqnarray
where $X_{n,\beta}$ and $\Xbar_{\beta,n}$ are linear
differential operators as in (\ref{eq:Toda-TauConstraint})
that give a $W_{1+\infty}$ symmetry of the Toda lattice
hierarchy. If $\beta = 1$, this is exactly the $W_{1+\infty}$
constraints that Dijkgraaf et al. \cite{bib:DMP} present
in the former half of their paper.  (Thus, as far as we
understand, they deal with two distinct cases in the
same paper --- the $\beta = 1$ case in the former half
and the $\beta =-1$ case in the latter half!)  Meanwhile,
if $\beta = -1,-2,\ldots$, the $W_{1+\infty}$ constraints
are of the form
\beqnarray
   n |\beta| t_{n|\beta|} \tau(t,\tbar,s)
   &=& \Xbar_{\beta,n}(\tbar,s,\rd_\tbar) \tau(t,\tbar,s),
                                       \nonumber \\
   n |\beta| \tbar_{n|\beta|} \tau(t,\tbar,s)
   &=& X_{\beta,n}(t,s,\rd_t) \tau(t,\tbar,s).
\eeqnarray

As a final remark, we would like to note that if $\hbar$ is
replaced by $-\hbar$ in the definition of $u_{ij}$, the string
equations become
\beqnarray
    L^\beta &=& \Bigl(  \frac{1}{\beta} \Mbar
      + \hbar \frac{\beta+1}{2\beta} + 1 \Bigr) \Lbar^\beta,
                                          \nonumber \\
    \Lbar^{-\beta} &=& \Bigl(  \frac{1}{\beta} M
      - \hbar \frac{\beta-1}{2\beta} + 1) L^{-\beta}.
\eeqnarray
In several papers, string equations of $c = 1$ or two-dimensional
topological strings are given in this form.

\section{Synthesis --- generalized string equations}

We now present our new family of special solutions and
associated string equations. These solutions have two
discrete parameters $(p,\pbar)$. The generalized
Kontsevich models and the compactified $c = 1$ string
theory can be reproduced by letting these parameters to
special values. In fact, there are two apparently different
constructions starting from the generating functions
$u_j(\lambda)$ and $\ubar_i(\mu)$, respectively, which
eventually lead to the same string equations. Unfortunately,
we have been unable to see if these two constructions
give the same solution.  We mostly present the first
construction based on $u_j(\lambda)$, and just briefly
mention the second one.

\subsection{Generating functions and string equations}

The generating functions $u_j(\lambda)$ of the new special solution
are given by
\beqnarray
    u_j(\lambda) &=&
      c(\lambda) \int d\mu \mu^{(\pbar - 1)/2 - j - 1}
      \exp \hbar^{-1} \Bigl( \lambda^p \mu^\pbar
      - \frac{\pbar}{p+\pbar} \mu^{p+\pbar} \Bigr),
                                            \nonumber \\
    c(\lambda) &=&
      \const \lambda^{(p-1)/2} \exp \hbar^{-1} \Bigl(
          \frac{-p}{p+\pbar} \lambda^{p+\pbar} \Bigr).
\eeqnarray
The case of $p + \pbar = 0$ is also included here by replacing
\beqn
    \frac{p}{p+\pbar} \lambda^{p+\pbar} \to p\log\lambda
\eeqn
in the limit as $\pbar \to - p$. In fact, $u_j(\lambda)$ then
becomes monomials of $\lambda$, and reduce to the generating
functions in the $c = 1$ string theory. The case of
$(p,\pbar) = (p,1)$, meanwhile, is nothing but the $p$-th
generalized Kontsevich model. The integrands of the above
integrals may have singularities at the origin, but this
apparent difficulty can be remedied by the same trick as
discussed in the previous cases. Furthermore, in much
the same way, one can derive the following relations for
general $(p,\pbar)$:
\beqnarray
    \lambda^p u_j(\lambda) &=&
      \hbar\Bigl( \frac{j+1}{\pbar} + \frac{\pbar-1}{2\pbar}
      \Bigr) u_{j+\pbar}(\lambda)
      + u_{j-p}(\lambda),
                                             \nonumber \\
    u_{j-\pbar}(\lambda) &=&
      \Bigl( \frac{\hbar}{p} \lambda^{1-p}\frac{\rd}{\rd\lambda}
      - \hbar \frac{p-1}{2p} \lambda^{-p} + \lambda^{\pbar}
      \Bigr) u_j(\lambda).
\eeqnarray

The solution of the Toda lattice hierarchy in question is
determined by asymptotic expansion of $u_j(\lambda)$. In fact,
whereas the case of $p + \pbar > 0$ is more or less parallel to
generalized Kontsevich models, the case of $p + \pbar < 0$ is
quite distinct --- we have to consider asymptotic expansion as
$\lambda \to +0$ rather than $\lambda \to +\infty$. To see this,
let us change the variable of integration from $\mu$ to a new
variable $z$ as:
\beqn
    \mu = \lambda z.
\eeqn
Then
\beqnarray
    u_j(\lambda)
    &=& \const \lambda^{(p+\pbar)/2 -j-1} \exp \hbar^{-1} \Bigl(
        \frac{-p}{p+\pbar} \lambda^{p+\pbar} \Bigr)
                                                \nonumber \\
    &&  \times
        \int dz z^{(\pbar-1)/2 -j-1} \exp (\hbar^{-1} \lambda^{p+\pbar})
        \Bigl( z^\pbar - \frac{\pbar}{p+\pbar} z^{p+\pbar} \Bigr).
                                  \label{eq:u(lambda)xi-integral}
\eeqnarray
Thus the saddle point analysis has to be done in a region
where $\hbar^{-1} \lambda^{p+\pbar} \to +\infty$. The most
dominant contribution comes from the saddle point $z = 1$,
which corresponds to $\mu = \lambda$ in the original integral.
Contributions from other saddle points (and from the endpoint
of the path of integration, if the path is cut off in a
neighborhood of the origin) are subdominant and negligible
in asymptotic expansion of $u_j(\lambda)$. The prefactor
on the right hand side of (\ref{eq:u(lambda)xi-integral})
cancels the leading quasi-classical contribution from the
integral, so that $u_j(\lambda)$ has asymptotic expansion
in negative integral powers of $\lambda^{p+\pbar}$.
In summary, we obtain the following result:

\begin{theorem}
(i) If $p + \pbar > 0$, $u_j(\lambda)$ has asymptotic expansion of
the form
\beqn
    u_j(\lambda) \sim \sum_{i=j}^\infty \lambda^{-i-1} u_{ij}
    \quad (\lambda \to +\infty).
\eeqn
In particular, $\bU$ is lower triangular.
\newline
(ii) If $p + \pbar < 0$, $u_j(\lambda)$ has asymptotic expansion of
the form
\beqn
    u_j(\lambda) \sim \sum_{i=-\infty}^j \lambda^{-i-1} u_{ij}
    \quad (\lambda \to +0).
\eeqn
In particular, $\bU$ is upper triangular.
\newline
(iii) In both cases, $u_{ij} = 0$
if $i - j \not\equiv 0 \ \mbox{mod} \ p+\pbar$.
\end{theorem}

We shall examine this asymptotic expansion in more detail in the
next subsection.  Anyway, the infinite matrix $\bU = \bU(0,0)$
determines a solution of the Toda lattice hierarchy. By the
triangular form of $\bU$, the $\tau$ function is ensured to
have a Miwa variable representation. One should however note
that one cannot use $u_j(\lambda)$ in the case of $p + \pbar < 0$.
If $p + \pbar < 0$ (so that $\bU$ is upper triangular),
it is $\ubar_i(\mu)$ that emerge in the Miwa variable
representation. Unfortunately, we have been unable to give
a closed expression to $\ubar_i(\mu)$ in the present setting.
[This should not be confused with the ``dual construction''
discussed later, which starts from an explicit construction of
$\ubar_i(\mu)$. In that case, the corresponding $u_j(\lambda)$
remains unknown in turn.] This is an incomplete aspect of the
present construction.

The aforementioned relations among the generating functions imply
the following linear relations among the coefficients $u_{ij}$
of the asymptotic expansion:
\beqnarray
    u_{i+p,j} &=&
    \hbar \Bigl( \frac{j+1}{\pbar} + \frac{\pbar-1}{2\pbar}
    \Bigr) u_{i,j+\pbar} + u_{i,j-p},
                                      \nonumber \\
    u_{i,j-\pbar} &=&
    \hbar \Bigl( - \frac{i}{p} + \frac{p-1}{2p}
    \Bigr) u_{i-p,j} + u_{i+\pbar,j}.
\eeqnarray
In terms of the matrix $\bU = (u_{ij})$,
\beqnarray
    \bLambda^p \bU &=&
    \bU \Bigl( \frac{\hbar}{\pbar} \bDelta \bLambda^{-\pbar}
    - \hbar \frac{\pbar-1}{2\pbar} \bLambda^{-\pbar}
    + \bLambda^p \Bigr),
                                                       \nonumber \\
    \bU \bLambda^{\pbar} &=&
    \Bigl( - \frac{\hbar}{p} \bDelta \bLambda^{-p}
    + \hbar \frac{p-1}{2p} \bLambda^{-p} + \bLambda^\pbar) \bU.
\eeqnarray

String equations can be readily derived from these relations:

\begin{theorem}
The Lax and Orlov-Shulman operators of this solution
obey the string equations
\beqnarray
    L^p &=&
    \frac{1}{\pbar} \Mbar \Lbar^{-\pbar}
    - \hbar \frac{\pbar-1}{2\pbar} \Lbar^{-\pbar}
    + \Lbar^p,
                                          \nonumber \\
    \Lbar^\pbar &=&
    - \frac{1}{p} M L^{-p}
    + \hbar \frac{p-1}{2p} L^{-p}
    + L^\pbar.
\eeqnarray
\end{theorem}

The string equations of generalized Kontsevich models and $c = 1$
strings can be reproduced by letting $(p,\pbar) = (p,1)$ and
$(p,\pbar) = (\beta,-\beta)$. The case of $(p,\pbar) = (1,-1)$
is also essentially the same as the string equations of the
two-matrix model ; the extra term $L$ and $\Lbar$ on the right hand
side can be absorbed into a shift of $s$.  (More precisely, it is a
``forced Toda lattice hierarchy'' confined to a semi-infinite lattice
\cite{bib:GKMvsToda} thar arises here.) The string equations for
general values of $(p,\pbar)$, too, relates ``coordinates'' $L$ and
$\Lbar$ to ``momenta'' $\Mbar \Lbar^{-\pbar}/\pbar$ and $M L^{-p}/p$.
In this respect, these string equations may be referred to as
``two-matrix model type'', as mentioned in Introduction.

\subsection{Systematic method of asymptotic expansion}

Let us examine the coefficients $u_{ij}$ of the asymptotic
expansion in more detail. We would eventually like to evaluate
them in an explicit form.  Such an explicit formula will be
useful in considering the relation between the present
construction and its ``dual'' form discussed later.

Since this issue can be treated in a more general context,
we now consider an integral of the general form
\beqn
    I(k) = \int dz g(z) e^{-kf(z)},
\eeqn
where $k$ is a large positive number, $f(z)$ and $g(z)$ are
holomorphic functions, and $f(z)$ is supposed to have no saddle
point other than a simple nondegenerate saddle point at
$z = z_0$ along the path of integration.  By the assumption,
$f(z)$ can be written
\beqn
    f(z) = f_0 + f_1 (z - z_0)^2 + O\Bigl((z - z_0)^3 \Bigr),
    \quad f_1 \not= 0,
\eeqn
in a neighborhood of $z_0$. Then $I(k)$ should have asymptotic
expansion of the following form as $k \to +\infty$:
\beqn
    I(k) \sim e^{-k f_0} \sum_{n=0}^\infty a_n k^{-n-(1/2)}.
\eeqn

The problem is how to evaluate these coefficients. From the field
theoretical standpoint, the most familiar method would be the
expansion into ``Feynman graphs''.  Namely, one separates $f(z)$
into the quadratic part and the rest, as shown above, treating
the latter as a ``perturbation'', and evaluate the final series of
integrals as Gaussian integrals.  This is actually a very
inefficient way, because $a_n$ is then given by a sum of various
Feynman graphs, and evaluating the sum is a hard task.  A more
efficient method is to resort to the Laplace method. A modernized
version of this method is presented by Berry and Howls
\cite{bib:BerryHowls} (whose main concern rather consists
in global issues like Stokes phenomena).  This method yields
a closed formula for $a_n$, as we now briefly present below.

A clue of the latter method is to rewrite the integral $I(k)$
by the new variable
\beqn
    y = \Bigl(f(z) - f_0\Bigr)^{1/2}
      = f_1^{1/2}(z - z_0) + O\Bigl( (z - z_0)^2 \Bigr).
\eeqn
Note that it is just a neighborhood of $z_0$ that actually
contributes to the above asymptotic expansion; ``modulo
subdominant terms'', one can replace the full path of
integration by a small piece in a neighborhood of $z_0$.
$y$ is a local coordinate in such a neighborhood of $z_0$.
The integral $I(k)$ can now be written into a Gaussian
integral:
\beqn
    I(k) = e^{-k f_0} \int dy \frac{dz}{dy} g(z) e^{-ky^2}
           + \mbox{ subdom. }
\eeqn
The subdominant terms ``subdom.'' are of course invisible
in asymptotic expansion, and negligible in the following
calculations. The Jacobian $dz/dy$ and the function $g(z)$
are both holomorphic functions of $y$ in a neighborhood of
$y = 0$.  Let us write the Taylor expansion of their
product as:
\beqn
    \frac{dz}{dy} g(z) = \sum_{n=0}^\infty h_n y^n.
\eeqn
The asymptotic expansion of the above integral can be obtained
by inserting this Taylor expansion,  formally interchanging
the order of summation and integration, and finally extending
the range of integration to the full real axis.  Final integrals
are to be evaluated by the familiar formula
\beqn
    \int_{-\infty}^\infty dy y^{2n} e^{-ky}
    = \Gamma\Bigl(n + \frac{1}{2}\Bigr) k^{-n-(1/2)}, \quad
    \int_{-\infty}^\infty dy y^{2n-1} e^{-ky}
    = 0.
\eeqn
Thus $a_n$ are given by
\beqn
    a_n = \Gamma\Bigl(n + \frac{1}{2}\Bigr) h_{2n}.
\eeqn
Furthermore, the Taylor coefficients $h_{2n}$ are given by
an integral of the form
\beqn
    h_{2n} = \frac{1}{2 \pi i} \oint_{y=0} dy \dfrac{dz}{dy}
             \dfrac{g(z)}{y^{2n+1}}
\eeqn
along a small contour around $y = 0$. This contour is mapped
onto a contour around $z = z_0$ on the $z$ plane, because
$y \to z$ is a coordinate transformation.  Thus the last
integral can be rewritten into a contour integral on the
$z$ plane:
\beqn
    h_{2n} = \frac{1}{2 \pi i} \oint_{z=z_0} dz
             \dfrac{g(z)}{\Big( f(z) - f_0 \Bigr)^{n+(1/2)}}.
\eeqn
In summary, we obtain the following formula for the coefficients
$a_n$:
\beqn
    a_n = \dfrac{\Gamma(n + \frac{1}{2})}{2 \pi i} \oint_{z=z_0} dz
          \dfrac{g(z)}{\Big( f(z) - f_0 \Bigr)^{n+(1/2)}}.
\eeqn

In the case of our $u_j(\lambda)$, $f(z)$ and $g(z)$ are given by
\beqn
    f(z) = \frac{\pbar}{p+\pbar} z^{p+\pbar} - z^\pbar, \quad
    g(z) = z^{(\pbar-1)/2 - j - 1}.
\eeqn
The relevant saddle point is at $z = 1$; the others are $p$-th
roots of unity and the origin, and all subdominant if we consider
the asymptotics as $\hbar^{-1}\lambda^{p+\pbar} \to +\infty$.
The above contour integrals are still hard to evaluate explicitly,
but anyway, we thus obtain a closed formula for the coefficients
$u_{ij}$.

\subsection{Dual construction}

A parallel construction is possible by starting from the
generating functions
\beqnarray
    \ubar_i(\mu) &=&
    \cbar(\mu) \int d\lambda \lambda^{(p-1)/2 + i}
    \exp \hbar^{-1} \Bigl( \lambda^p \mu^\pbar
    - \frac{p}{p+\pbar} \lambda^{p+\pbar} \Bigr),
                                            \nonumber \\
    \cbar(\mu) &=&
    \const \mu^{(\pbar-1)/2} \exp \hbar^{-1} \Bigl(
    \frac{-\pbar}{p+\pbar} \mu^{p+\pbar} \Bigr).
\eeqnarray
Previous calculations on $u_j(\lambda)$ can be repeated
in a fully parallel manner. The generating functions
$\ubar_i(\mu)$ turn out to have asymptotic expansion of
the form
\beqn
    \ubar_i(\mu) \sim \sum_{j=-\infty}^i \ubar_{ij} \mu^j
    \quad (\mu \to +\infty)
\eeqn
if $p + \pbar > 0$, and
\beqn
    \ubar_i(\mu) \sim \sum_{j=i}^\infty \ubar_{ij} \mu^j
    \quad (\mu \to +0)
\eeqn
if $p + \pbar < 0$. (The case of $p + \pbar = 0 $ again reduces
to the $c = 1$ string theory.)  The coefficients $\ubar_{ij}$
satisfy exactly the same linear relations as the coefficients
$u_{ij}$ in the previous construction.  In other words,
this ``dual'' construction leads to a solution of the same
string equations.

We have been, however, unable to prove (or disprove) that these
two constructions in fact give the same solution.  An obstruction
is the fact that the linear relations among $u_{ij}$ appear to allow
more ambiguities than the overall rescaling $u_{ij} \to \const u_{ij}$.
Nevertheless, it seems likely that the above solutions are rather
special and coincide.  The contour integral representation in the
previous subsection will be useful in pursuing this issue.

\subsection{Concluding remark of this section}

The contents of this section is inspired by the work of Kharchev
and Marshakov \cite{bib:KharMarPQ} on the $(p,q)$ duality in
$c < 1$ strings. Our $\bU$ is a matrix representation of their
Fourier-Laplace integral operator that gives the $(p,q)$ duality.
Namely, if $V_{pq}$ denotes the point representing the $(p,q)$
model in the Sato Grassmannian, $\bU$ acts on the Sato Grassmannian
and interchanges the $(p,q)$ and $(q,p)$ models:
\beqn
    \bU V_{qp} = V_{pq}.
\eeqn
Note that everything is formulated in the language of the KP
hierarchy.

Although closely related with their work, our usage of $\bU$ is
substantially different.  We rather interpret $\bU$ as a $GL(\infty)$
element that determines a solution of the Toda lattice hierarchy.
Only in the case of $q = 1$, their work and ours are directly related
via generalized Kontsevich models.  Recall, however, that even in
that case, our interpretation of generalized Kontsevich models is
slightly different from the ITEP-Lebedev group \cite{bib:GKMvsToda}.
We have pointed out that the generalized Kontsevich models with
``negative times'' give a solutions on the full (bi-infinite)
lattice rather than a half (semi-infinite) lattice.

\section{Conclusion and discussion}

We have constructed a new family of special solutions of the
Toda lattice hierarchy, and derived string equations of these
solutions. These solutions have two discrete parameters
$(p,\pbar)$, and include already known solutions as follows:
\begin{itemize}
\item
    $(p,\pbar) = (p,1)$, $p > 0$ --- the $p$-th generaized
    model with ``negative times'' $\tbar$ and a ``discrete time'' $s$.
\item
    $(p,\pbar) = (\beta,-\beta)$, $\beta > 0$ --- the compactified
    $c = 1$ string theory. $\beta = 1$ corresponds to the self-dual
    radius case.
\item
    $(p,\pbar) = (-1, 1)$ --- related to the Penner model. This
    case may be interpreted as the generalized Kontsevich model
    at $p = -1$, and also as the $c = 1$ string theory ``at
    $\beta = -1$''.
\end{itemize}

In the course of reviewing known examples, we have also posed
several questions on the relation \cite{bib:DMP,bib:Ozetal}
between the generalized Kontsevich models and the $c = 1$
string theory. In particular, as opposed to the hypothesis
postulated in the literature \cite{bib:Ozetal}, the naive
extrapolation of the $p$-th  generalized Kontsevich model
to $p = -1$ does not give the $c = 1$ string theory at
self-dual radius. The $p = -1$ limit is rather related
to the Penner model \cite{bib:Penner}. Actually,
these three models are mutually distinct and obey
different string equations. Thus, although there
is a lot of evidence \cite{bib:Witten} that the
$A_{k+1}$ model at $k = -3$ is anyhow related to
$c = 1$ (or two-dimensional) strings, the approach
to $c = 1$ by the generalized Kontsevich models seems
to be problematical.

A natural extrapolation of the $c = 1$ string theory
at self-dual radius will be given by the case of $p = 1$
and $\pbar \le -1$.  This model is ensured to reduced
to the $c = 1$ model by letting $\pbar \to -1$.
Furthermore, remarkably, this model has a matrix integral
representation of Kontsevich type, as follows. The matrix
$\bU = \bU(0,0)$ of this model is upper triangular;
the $\tau$ function has a Miwa variable representation
by the Miwa transformation
\beqn
    \tbar_n = - \frac{\hbar}{n} \sum_{i=1}^N \mu_i^{-n}
            = - \frac{\hbar}{n} \Tr M^{-n}.
\eeqn
This Miwa variable representation can be rewritten into a
matrix integral of the form
\beqnarray
    &&  \tau(t,\tbar,\hbar(N-1))
                               \nonumber \\
    &=& \Cbar(M) \int d\Lambda \exp \hbar^{-1} \Tr \Big(
        \Lambda M^\pbar - \frac{1}{1+\pbar} \Lambda^{1+\pbar}
        + \sum_{n=1}^\infty t_n \Lambda^n \Bigr),
\eeqnarray
where we have considered the case of $s = \hbar(N-1)$
for simplicity. (This expression can be readily extended
to general $s$.)

Unlike the ordinary generalized Kontsevich models,
the Kontsevich potential $\Lambda^{1+\pbar}/(1 + \pbar)$
in this matrix integral is a negative power of $\Lambda$.
Therefore the integral should be suitably interpreted (e.g.,
by suitable regularization to avoid essential singularities
as $\det \Lambda \to 0$).  The $\tau$ function itself, as a
function (or formal power series) of $(t,\tbar)$, is however
independent of such a detail of regularization, and only
determined by the contribution from the saddle point
$\Lambda = M$. The above matrix integral representation
might lead to a topological interpretation of this model
by means of ``fat graph'' expansion.

Another important difference from the ordinary generalized
Kontsevich models is that the string equations of this model
never reduces to string equations in the KP hierarchy.
In the case of generalized Kontsevich models, the
``negative times'' are coupling constants of unphysical
states (``anti-states'' in the language of Montano and
Rivlis \cite{bib:MontanoRivlis}), which decouple from
the theory as $\tbar \to 0$, and the string equations turn
into string equations in the KP hierarchy in this limit.
This does not occur in the case of $p = 1$ and $\pbar \le -1$.
In other words, the model persists to be ``$c = 1$'' in
the whole space of the coupling constants.

We, however, do not know what physical interpretation the
string equations actually have for general values of $(p,\pbar)$
including the above case of $p = 1$ and $\pbar \le -1$.
Do they all correspond to a physical model (hopefully, of strings) ?
A first step towards such deeper understanding will be to show a
connection with moduli spaces of Riemann surfaces. The works of
Montano and Rivlis \cite{bib:MontanoRivlis} and Lavi et al.
\cite{bib:Ozetal} will provide useful ideas in that direction.

\begin{acknowledgements}

The author is very grateful to Toshio Nakatsu, Takahiro Shiota
and Takashi Takebe for many comments and encouragement. He is
also indebted to Chris Howls for ideas on asymptotic analysis.
This work is partly supported by the Grants-in-Aid for Scientific
Research, Priority Area 231 ``Infinite Analysis'', the Ministry
of Education, Science, Culture, Japan.  Part of this work was
completed while participating in the research program ``Exponential
Asymptotics'' at the Isaac Newton Institute, the University of
Cambridge.
\end{acknowledgements}

\begin{noteadded}

After completed this paper, the author learned that
the same issue as we have considered on $c = 1$ strings
was studied  by Imbimbo and Mukhi in a recent preprint
\cite{bib:ImbimboMukhi}. They noted the inconsistency
between the Kontsevich-type representation of Dijkgraaf
et al. \cite{bib:DMP} and the $W_{1+\infty}$ constraints,
and proposed an alternative matrix model. Our matrix
integral representation of the $c = 1$ partition
function in Section 4 seems to be essentially the
same as this model. The author would like to thank
Sunil Mukhi for pointing out the presence of their paper.
\end{noteadded}



\begin{thebibliography}{99}

\bibitem{bib:AvM}
M. Adler and P. van Moerbeke,
A matrix integral solution to two-dimensional $W_p$-gravity,
Commun. Math. Phys. {\bf 147} (1992), 25-56.

\bibitem{bib:ASvM}
M. Adler, T. Shiota  and P. van Moerbeke,
A Lax Representation for the Vertex Operator
and the central extension,
Commun. Math. Phys. (to appear).

\bibitem{bib:BerryHowls}
M.V. Berry and C.J. Howls,
Hyperasymptotics for integrals with saddles,
Proc. R. Soc. London {\bf A434} (1991), 657-675.

\bibitem{bib:BXMMM}
L. Bonora and C.S. Xiong,
Matrix models without scaling limit,
Int. J. Mod. Phys. {\bf A8} (1993), 2973-2992;
Multimatrix models without continuum limit,
Nucl. Phys. {\bf B405} (1993), 191-227.

\bibitem{bib:BX2Dstring}
L. Bonora  and C.S. Xiong,
Two-matrix model and $c=1$ string theory,
Phys. Lett. {\bf B347} (1995), 41-48;
Extended Toda lattice hierarchy, extended two-matrix model
and $c = 1$ string theory,
Nucl. Phys. {\bf B434} (1995), 408-444.

\bibitem{bib:Dijkgraaf}
R. Dijkgraaf,
Intersection theory, integrable hierarchies, and
topological field theory,
in {\it New Symmetry Principles in Quantum Field Theory\/},
NATO ASI Carg\`ese 1991 (Plenum, 1991).

\bibitem{bib:DMP}
R. Dijkgraaf, G. Moore and R. Plesser,
The partition function of 2D string theory,
Nucl. Phys. B394 (1991), 356-382.

\bibitem{bib:Douglas}
M. Douglas,
Strings in less than one-dimension and
the generalized K-dV hierarchies,
Phys. Lett. {\bf 238B}  (1990) 176-180.

\bibitem{bib:EguchiHoriYang}
T. Eguchi, K. Hori and S.-K. Yang,
Topological sigma models and large $N$ matrix integral,
UT-700, hep-th/9503017.

\bibitem{bib:EK2Dstring}
T. Eguchi and H. Kanno,
Toda lattice hierarchy and the topological description
of $c = 1$ string theory,
Phys. Lett. {\bf B331} (1994), 330-334.

\bibitem{bib:EguchiYang}
T. Eguchi and S.-K. Yang,
The topological $CP^1$ model and the large $N$ matrix integral,
Mod. Phys. Lett. {\bf A9} (1994), 2893-2902.

\bibitem{bib:FKN}
M. Fukuma, H. Kawai and R. Nakayama,
Infinite dimensional Grassmannian structure of
two dimensional string theory,
Commun. Math. Phys. {\bf 143} (1991), 371-403.

\bibitem{bib:MMMToda}
A. Gerasimov, A. Marshakov, A. Mironov, A. Morozov and A. Orlov,
Matrix models of 2D gravity and Toda theory,
Nucl. Phys. {\bf B357} (1991), 565-618.
\newline
E.J. Martinec,
On the origin of integrability in matrix models,
Commun. Math. Phys. {\bf 138} (1991), 437-450.
\newline
L. Alvarez-Gaum\'e, C. Gomez and J. Lacki,
Integrability in random matrix models,
Phys. Lett. {\bf B253} (1991), 56-62.

\bibitem{bib:GhoshalMukhi}
D. Ghoshal and S. Mukhi,
Topological Landau-Ginzburg model of two-dimensional
string theory,
Nucl. Phy. {\bf B425} (1994), 173-190.

\bibitem{bib:Goeree}
J. Goeree,
W constraints in 2d quantum gravity,
Nucl. Phys. {\bf B358} (1991), 737-757.

\bibitem{bib:HOP}
A. Hanany, Y. Oz and R. Plesser,
Topological Landau-Ginzburg formulation and integrable
structure of 2d string theory,
Nucl. Phys. {\bf B425} (1994), 150-172.

\bibitem{bib:ImbimboMukhi}
C. Imbimbo and S. Mukhi,
The toplogical matrix model of $c = 1$ string,
hep-th/9505127.

\bibitem{bib:KannoOhta}
H. Kanno and Y. Ohta,
Topological strings with scaling violation and
Toda lattice hierarchy,
Nucl. Phys. {\bf B442} (1995), 179-204.

\bibitem{bib:GKMvsToda}
S. Kharchev, A. Marshakov, A. Mironov and A. Morozov,
Generalized Kontsevich model versus Toda hierarchy and
discrete matrix models,
Nucl. Phys. {\bf B397} (1993), 339-378.

\bibitem{bib:KharMarPQ}
S. Kharchev and A. Marshakov,
Topological versus non-topological theories and $p-q$
duality in matrix models,
presented at Rome String Theory Workshop 1992,
FIAN/TD-15/92, hep-th/9210072;
On $p-q$ duality and explicit solutions in $c \le 1$
2d gravity models,
Int. J. Mod. Phys. {\bf A10} (1995), 1219-1239.

\bibitem{bib:Kontsevich}
M. Kontsevich,
Intersection theory on the moduli space of curves and
the matrix Airy function,
Commun. Math. Phys. {\bf 147} (1992), 1-23.

\bibitem{bib:Ozetal}
Y. Lavi, Y. Oz and J. Sonnenschein,
$(1,q=-1)$ model as a topological description of 2d string theory,
Nucl. Phys. {\bf B431} (1994), 223-227.
\newline
A. Hanany and Y. Oz,
$c = 1$ discrete states correlators via $W_{1+\infty}$ constraints,
Phys. Lett. {\bf B347} (1995), 255-259.
\newline
Y. Oz,
On topological 2D string and intersection theory,
TAUP-2234-95, hep-th/9502058.

\bibitem{bib:Marshakov}
A. Marshakov,
On the string field theory for $c \le 1$,
in {\it Pathways to Fundamental Theories\/}
(World Scientific, Singapore, 1993).

\bibitem{bib:MontanoRivlis}
D. Montano and G. Rivlis,
Solving topological 2D quantum gravity using Ward identities,
Nucl. Phys. {\bf B404} (1993), 483-516.

\bibitem{bib:Morozov}
A. Morozov,
Integrability and matrix models,
ITEP-M2/93, ITFA 93-10, hep-th/9303139,
and references cited therein.

\bibitem{bib:Mulase}
M. Mulase,
Complete integrability of the Kadomtsev-Petviashvili equation,
Advances in Math. {\bf 54} (1984), 57-66;
Solvability of the super KP equation and
a generalization of the Birkhoff decomposition,
Invent. Math. {\bf 92} (1988), 1-46.

\bibitem{bib:Nakatsu}
T. Nakatsu,
On the string equation at $c = 1$,
Mod. Phys. Lett. {\bf A9} (1994), 3313-3324.

\bibitem{bib:NTTceq1}
T. Nakatsu, K. Takasaki and S. Tsujimaru,
Quantum and classical aspects of deformed $c = 1$ strings,
INS-rep.-1087, KUCP-0077, hep-th/9501038,
Nucl. Phys. {\bf B} (to appear).

\bibitem{bib:Orlovetal}
A.Yu. Orlov and E.I. Schulman,
Additional symmetries for integrable equations
and conformal algebra representation,
Lett. Math. Phys. {\bf 12} (1986), 171-179.
\newline
A.Yu. Orlov,
Vertex operators, $\bar{\partial}$-problems, symmetries,
variational indentities and Hamiltonian formalism for
$2+1$ integrable systems,
in {\it Plasma Theory and Nonlinear and
Turbulent Processes in Physics\/}
(World Scientific, Singapore, 1988).
\newline
P.G. Grinevich and A.Yu. Orlov,
Virasoro action on Riemann surfaces, Grassmannians,
$\det\bar{\partial}_j$ and Segal Wilson $\tau$ function,
in {\it Problems of Modern Quantum Field Theory\/}
(Springer-Verlag, 1989).

\bibitem{bib:Penner}
R.C. Penner,
Perturbative series and the moduli space of Riemann surfaces,
J. Diff. Geom. {\bf 27} (1988), 35-53.

\bibitem{bib:KP}
M. Sato  and  Y. Sato,
Soliton equations as dynamical systems on infinite dimensional
Grassmann manifold,
in {\it Nonlinear Partial Differential Equations in Applied Science;
Proceedings of the U.S.-Japan Seminar, Tokyo, 1982},
Lect. Notes in Num. Anal. {\bf 5} (1982), 259-271.
\newline
M. Sato and M. Noumi,
Soliton equations and the universal Grassmann manifolds,
Sophia Univ. Kokyuroku in Math. {\bf 18} (1984),
in Japanese.
\newline
E. Date, M. Kashiwara, M. Jimbo and  T. Miwa,
Transformation groups for soliton equations,
in {\it Nonlinear Integrable Systems --
Classical Theory and Quantum Theory}
(World Scientific, Singapore, 1983), 39-119.
\newline
G. Segal  and  G. Wilson,
Loop groups and equations of KdV type,
IHES Publ. Math. {\bf 63} (1985) 1-64.

\bibitem{bib:KacSchwarz}
A. Schwarz,
On solutions to the string equations,
Mod. Phys. Lett. {\bf A6} (1991), 2713-2726.
\newline
V. Kac and A.Schwarz,
Geometric interpretation of partition function of 2D gravity,
Phys. Lett. {\bf B257} (1991), 329-334.

\bibitem{bib:TodaIVP}
K. Takasaki,
Initial value problem for the Toda lattice hierarchy,
in {\it Group Representations and Systems of Differential Equations},
K. Okamoto ed., Advanced Studies in Pure Math. {\bf 4}
(North-Holland/Kinokuniya 1984).
\newline
T. Takebe,
Representation theoretical meaning of the initial value problem
for the Toda lattice hierarchy I,
Lett. Math. Phys. {\bf 21} (1991), 77-84;
ditto II,
Publ. RIMS, Kyoto Univ., {\bf 27} (1991), 491-503.

\bibitem{bib:T2Dstring}
K. Takasaki,
Dispersionless Toda hierarchy and two-dimensional string theory,
Commun. Math. Phys. {\bf 170} (1995), 101-116.

\bibitem{bib:TTToda}
K. Takasaki and T. Takebe,
SDiff(2) Toda equation -- hierarchy, tau function and symmetries,
Lett. Math. Phys. {\bf 23} (1991), 205-214;
Quasi-classical limit of Toda hierarchy and W-infinity symmetries,
Lett. Math. Phys. {\bf 28} (1993), 165-176.

\bibitem{bib:TTreview}
K. Takasaki and  T. Takebe,
Integrable hierarchies and dispersionless limit,
UTMS 94-35, hep-th/9405096,  Rev. Math. Phys. (to appear),
and references cited therein.

\bibitem{bib:UTToda}
K. Ueno  and K. Takasaki,
Toda lattice hierarchy,
in {\it Group Representations and Systems of Differential Equations},
K. Okamoto ed., Advanced Studies in Pure Math. {\bf 4}
(North-Holland/Kinokuniya 1984).

\bibitem{bib:vandeLeur}
J. van de Leur,
KdV type hierarchies, the string equations and
$W_{1+\infty}$ constraints,
Utrecht-843, hep-th/9403080;
The $W_{1+\infty}(gl_s)$ symmetries of
the $s$ component KP hierarchy,
hep-th/9411069.

\bibitem{bib:Witten}
E. Witten,
The $N$ matrix model and gauged WZW models,
Nucl. Phys. {\bf B371} (1992), 191-245;
Algebraic geometry associated with matrix models
of two dimensional gravity,
Nucl. Phys. {\bf B377} (1992), 55-112.

\bibitem{bib:Yoneya}
T. Yoneya,
Toward a canonical formalism of non-perturbative
two-dimensional gravity,
Commun. Math. Phys. {\bf 144} (1992), 623-639.

\end{thebibliography}
\end{document}